\documentclass[11pt]{article}

\usepackage{mathptmx}
\usepackage{amsmath,amssymb}
\usepackage{color}

\numberwithin{equation}{section}
\newcommand{\e}{\epsilon}
\newcommand{\p}{\partial}
\newcommand{\nn}{\nonumber}
\newcommand{\R}{\mathcal{R}}
\newcommand{\A}{\mathcal{A}}
\newcommand{\B}{\mathcal{B}}
\newcommand{\D}{\mathcal{D}}
\newcommand{\LL}{\Lambda}
\newcommand{\Lie}{\mathrm{Lie}}
\newcommand{\ad}{\mathrm{ad}}

\newtheorem{thm}{Theorem}[section]
\newtheorem{prp}[thm]{Proposition}
\newtheorem{lem}[thm]{Lemma}
\newtheorem{dfn}[thm]{Definition}
\newtheorem{cor}[thm]{Corollary}
\newtheorem{emp}[thm]{Example}
\newtheorem{rmk}[thm]{Remark}
\newenvironment{prf}{\noindent {\it Proof }}{\hfill $\square$}

\begin{document}

\title{On Properties of Hamiltonian Structures for a Class of Evolutionary PDEs }
\author{Si-Qi Liu\thanks{liusq@mail.tsinghua.edu.cn}
\quad Chao-Zhong Wu\thanks{wucz05@mails.tsinghua.edu.cn}
\quad Youjin Zhang\thanks{youjin@mail.tsinghua.edu.cn} \\
{\small Department of Mathematical Sciences, Tsinghua University, }\\
{\small Beijing, P. R. China}}
\date{}
\maketitle

\begin{abstract}
In \cite{LZ2} it is proved that for certain class of
perturbations of the hyperbolic equation $u_t=f(u) u_x$, there exist
changes of coordinate, called quasi-Miura transformations, that
reduce the perturbed equations to the unperturbed one. We prove in
the present paper that if in addition the perturbed equations
possess Hamiltonian structures of certain type, the same
quasi-Miura transformations also reduce the Hamiltonian structures
to their leading terms. By applying this result, we obtain a
criterion of the existence of Hamiltonian structures for a class of
scalar evolutionary PDEs and an algorithm to find
out the Hamiltonian structures.
\end{abstract}

\section{Introduction}

We consider in this paper the following class of generalized scalar evolutionary PDEs of the unknown function $u=u(x,t)$
\begin{equation}
u_t=f(u)u_x+\sum_{k=1}^\infty \e^k F_k(u; u_x, \cdots, u^{(k+1)}), \quad f'(u)\ne 0, \label{pde}
\end{equation}
where $u_t=\p_tu,\ u_x=\p_xu, \ u^{(\ell)}=\p_x^{\ell}u$.
The functions $F_k$ are assumed to be polynomials of $u_x, \dots, u^{(k+1)}$ with coefficients
depending smoothly on $u$,
and they are homogeneous of degree $k+1$ under the following assignment of degrees:
\begin{equation}
\deg u^{(\ell)}=\ell,\ \ell\ge1;\quad \deg h(u)=0 \mbox{ for a smooth function $h(u)$}. \label{deg}
\end{equation}
Such functions are called homogeneous
differential polynomials.

In the case when the right hand side of \eqref{pde} truncates, the equation is a usual evolutionary PDE.
In generally, we allow the the right hand side of \eqref{pde} to be non-truncated.
Such equations arise, for example, when we consider an evolutionary PDE of the form
\begin{equation}
u_t=K(u; u_x, \cdots, u^{(N)}) \label{k-pde}
\end{equation}
with a function $K$ being analytic at $u_x=\cdots=u^{(N)}=0$ and $K(u; 0, \cdots, 0)=0$.
We can make the rescaling $x\mapsto \e\,x,\ t\mapsto \e\,t$,
and expand the the right hand side of the equation \eqref{k-pde} into a power series of $\e$,
the resulting equation is of the form \eqref{pde}. Equations of the form \eqref{pde} also appear
when we consider PDEs of the following type
\begin{equation}
u_t-u_{xxt}=K(u; u_x, \cdots, u^{(N)}), \label{ch-pde}
\end{equation}
where the function $K$ is given as above. An important example of such class of equations
is the Camassa - Holm equation (see Example \ref{ep2} below) which is an integrable nonlinear
PDE describing shallow water waves \cite{CH1, CH2, FF}.
We can perform the same rescaling and rewrite the equation \eqref{ch-pde} as follows:
\[u_t=(1-\e^2\p_x^2)^{-1}\frac{K(u;\e u_x,\dots)}{\e}=\sum_{k=0}^\infty \e^{2k}\p_x^{2k}\frac{K(u;\e u_x,\dots)}{\e},\]
the right hand side of the above equation  can also be expanded into the form \eqref{pde}. Here the Taylor expansion converges
in the formal power
series topology.

In this paper we consider properties of the Hamiltonian structures
for equations of the form \eqref{pde}. Note that equation
\eqref{pde} is a perturbation of the hyperbolic equation
\begin{equation}
v_t=f(v)v_x,\quad f'(v)\ne 0. \label{pde0}
\end{equation}
This equation possesses infinitely many Hamiltonian structures. In fact, for any smooth function $g(v)$,
there is a Poisson bracket
on the space of functionals of $v(x)$ which is defined by
\begin{equation}
\{v(x), v(y)\}_g=g(v)\delta'(x-y)+\frac12 g'(v)v_x\delta(x-y), \label{ham0}
\end{equation}
here $v(x), v(y)$ are regarded as distributions. Then equation \eqref{pde0} can be expressed as
a Hamiltonian system
\[v_t=\{v(x), H_g\}_g,\]
where the functional $H_g$ is defined by
\begin{equation}\label{hham0}
H_g=\int h(v)\,dx, \mbox{ with $h$ satisfying } g\,h''+\frac12g'h'=f.
\end{equation}

We call the equation \eqref{pde} a {\em Hamiltonian system} or a
{\em{Hamiltonian perturbation}} of the equation \eqref{pde0} if
it possesses a Hamiltonian
structure
\begin{equation}
u_t=\{u(x), H\},\nn
\end{equation}
where the Poisson bracket and the Hamiltonian have the forms
\begin{align}
\{u(x), u(y)\}=&g(u)\delta'(x-y)+\frac12 g'(u)u_x\delta(x-y)\nn\\
&\quad+\sum_{k=1}^\infty \e^k\sum_{l=0}^{k+1}A_{k,l}(u;u_x,\cdots, u^{(l)})\delta^{(k+1-l)}(x-y), \label{ham}\\
&\hskip -2cm
H=\int\left(h(u)+\sum_{l=1}^\infty \e^l B_l(u;u_x,\cdots, u^{(l)})\right)\,dx. \label{hham}
\end{align}
Here $A_{k,l}$, $B_l$ are homogenous differential polynomials of degree $l$.

Studies of Hamiltonian perturbations of general hyperbolic systems were initiated by Dubrovin and the authors
of the present paper in \cite{DLZ}. It was proved there that any bihamiltonian perturbation of a hyperbolic
system which possesses a semisimple bihamiltonian structure is quasi-trivial, this means
that the perturbation terms can be eliminated by a change of the dependent variables, called a quasi-Miura
transformation, of the system. In \cite{Du}, Dubrovin studied the Hamiltonians perturbation of \eqref{pde0},
and proved certain universality of behavior of solutions of the perturbed equation \eqref{pde} near the point
of gradient catastrophe.
He also proved that the property of quasi-triviality still holds true at the approximation up to $\e^4$
for Hamiltonian perturbations of \eqref{pde0}, and it plays an important role in obtaining the main result of \cite{Du}.
A further study of the quasi-triviality property of equations of the form \eqref{pde} was given in \cite{LZ2},
where the following theorem is given on the validity of this property without the assumption of
existence of Hamiltonian structures.

\begin{thm}[Quasi-Triviality Theorem]
For any evolutionary PDE of the form \eqref{pde}, there exists a change of the dependent variable
called a quasi-Miura transformation
\begin{equation}
v=u+\sum_{k=1}^\infty \e^k \frac1{u_x^{L_k}}\sum_{m=0}^{M_k}Y_{k,m}(u; u_x, \cdots, u^{(N_k)})(\log u_x)^m \label{qt1}
\end{equation}
which reduces the equation \eqref{pde} to its leading term \eqref{pde0}. Here $L_k, M_k, N_k$ are certain integers that only
depend on $k$, and $Y_{k,m}$ are homogenous differential polynomials of degree $L_k+k$.
\end{thm}
The quasi-Miura transformation of the above theorem is also called the {\em reducing transformation} for \eqref{pde}.
In \cite{LZ2} this theorem is also applied to obtain a criterion of integrability for the class of
equations of the form \eqref{pde}.

Now given a Hamiltonian perturbation \eqref{pde},
we consider in this paper the relationship between the Hamiltonian structure and its reducing transformation.
The main result is the following:
\begin{thm}[Main Theorem]
For any Hamiltonian perturbation \eqref{pde} of the equation
\eqref{pde0} with a Hamiltonian structure of the form \eqref{ham}, \eqref{hham},
the reducing transformation \eqref{qt1} does not depend on $\log
u_x$, and it also reduces the Hamiltonian
structure \eqref{ham}, \eqref{hham} to its leading term \eqref{ham0}, \eqref{hham0}.
\end{thm}

By inverting the quasi-Miura transformation, the theorem implies
that any Hamiltonian structure of \eqref{pde} comes from a
Hamiltonian structure of the form \eqref{pde0}. This fact provides a
method to determine whether an equation of
the form \eqref{pde} possesses Hamiltonian structures, and if so how
to find them.

The paper is organized as follows: in the
next section, we give a proof of the main theorem; in Section 3, we
explain in detail the application of the Main Theorem to determine
the existence of Hamiltonian structures for evolutionary PDEs of the
form \eqref{pde}, and to find out the Hamiltonian structures. In the
final section we give some concluding remarks.

\section{Proof of the main theorem}

We first recall some useful notations from \cite{LZ2}. Denote $u_{\ell}=u^{(\ell)}\ (\ell=0, 1, 2, \cdots)$,
and define the ring of differential polynomials by $\R=C^\infty(u_0)[u_1, u_2, \cdots]$.
It is a graded ring with degrees given by \eqref{deg}. Let
$\A$ be its formal completion
\[\A=\{f=\sum_{i=0}^\infty \e^i f_i|f_i\in\R,\ \deg f_i=i\}.\]
Elements of $\A$ are also called differential polynomials.

For any $X\in\A$, we can define a
derivation on $\A$
\[\hat{X}=\sum_{i=0}^\infty \p_x^i X \frac{\p}{\p u_i}.\]
The map $\A\to\mathrm{Der}\A,\ X\mapsto\hat{X}$ induces the following Lie bracket on $\A$:
\[[X,Y]=\hat{X}(Y)-\hat{Y}(X),\ \forall\,X, Y\in\A,\]
and $\A$ becomes a module of this Lie algebra.
The Lie algebra $\A$ contains a subalgebra
\[\B=\{f=\sum_{i=1}^\infty \e^i f_i|f_i\in\R, \deg f_i=i\},\]
each element $X\in\B$ defines an evolutionary PDE of the form \eqref{pde}
\[u_t=\e^{-1} X=f(u)u_x+\e\,f_2+\e^2\,f_3+\cdots,\]
(without the assumption $f'(u)\ne0$ in general), so we will also  call $X$ an evolutionary PDE.
If in additional $f'(u)\ne0$, we say that the above PDE is generic.
Let $X\in\B$ be an evolutionary PDE, $h\in\A$ be a differential polynomial.
If there exists a $\sigma\in\A$ such that $\hat{X}(h)=\p_x \sigma$,
then $h$ is called a conserved density of $X$.

Consider the quotient space
\begin{equation}\label{zh-1}
\LL=\A/(\p_x\A\oplus\mathbb{R}),
\end{equation}
the coset corresponding to $f\in\A$ is denoted by $\int f\,dx$, it
is called a local functional defined by the density $f$. Note that
$\int\cdot\,dx$ is just a formal notation, it doesn't mean
integration here. It's easy to see that $\LL$ is also
an $\A$-module (as a module of
Lie algebra), the action of $X$ on $F=\int f\,dx$ is
given in the following natural way:
\begin{equation}
X.F=\int \hat{X}(f)\,dx=\int X\,\frac{\delta F}{\delta u}\,dx, \label{xf}
\end{equation}
where
\[\frac{\delta F}{\delta u}=\sum_{i=0}^\infty(-\p_x)^i \frac{\p f}{\p u^i}\]
is the variational derivative. In terms of these notations, $h\in\A$
is a conserved density of $X$ if and only if the local functional
$H=\int h\,dx$ satisfies $X.H=0$. Local
functionals of this type are called the conserved quantities of $X$.

Let us introduce the algebra of differential operators on $\A$ as follows:
\[\D=\{P=\sum_{k=0}^\infty\e^k\sum_{s=0}^{k+1}P_{k,s}\p_x^{k+1-s}|P_{k,s}\in\R,\ \deg P_{k,s}=s\}.\]
It is a graded algebra under the definition
\eqref{deg} and $\deg \p_x=1$. A differential operator $P\in\D$ is
called a Hamiltonian operator if the bilinear operation
\begin{equation}
\label{fglie} \{\cdot,\cdot\}_P:\LL\times\LL\to\LL,\quad
(F,G)\mapsto \int \frac{\delta F}{\delta u}P\left(\frac{\delta
G}{\delta u}\right)\,dx\end{equation} forms a Lie bracket. Note that
every Hamiltonian structure of the form
\eqref{ham} corresponds to a unique
Hamiltonian operator
\[
P=g(u)\p_x+\frac12 g'(u)u_x+ \sum_{k=1}^\infty
\e^k\sum_{l=0}^{k+1}A_{k,l}(u;u_x,\cdots, u^{(l)})\p_x^{k+1-l}
\]
and vice versa.
\begin{emp}
Let $P$ be a Hamiltonian operator of the form
\[P=P_1+\e^{k-1}\,P_k+O(\e^k),\quad k\ge 2,\]
it is easy to see that the leading term
$P_1=P_{10}\p_x+P_{11}$ is also a Hamiltonian operator. The
skew-symmetry condition of the Lie bracket
\eqref{fglie} implies that $P_1$ must be of the form
\begin{equation}
P_1=\phi(u)\p_x+\frac12\phi'(u)u_x. \label{p0}
\end{equation}
One can verify that a differential operator of the form \eqref{p0} is automatically a Hamiltonian operator.
The Jacobi identity for $\{\cdot, \cdot\}_P$ implies that the equality
\begin{align*}
&\{\{F,G\}_{P_k},H\}_{P_1}+\{\{F,G\}_{P_1},H\}_{P_k}+\{\{G,H\}_{P_k},F\}_{P_1}+\\
&\{\{G,H\}_{P_1},F\}_{P_k}+\{\{H,F\}_{P_k},G\}_{P_1}+\{\{H,F\}_{P_1},G\}_{P_k}=0
\end{align*}
holds true for any local functionals $F, G, H\in\LL$. The
differential operator $\epsilon^{k-1}P_k$ satisfying the above condition is called an infinitesimal
deformation of the Hamiltonian operator $P_1$.
\end{emp}

\begin{dfn}
An evolutionary PDE $X\in\B$ is called to
be Hamiltonian if there exists a Hamiltonian operator
$P$ and a local functional $H$ such that
\begin{equation}
X=P.H=P\left(\frac{\delta H}{\delta u}\right). \label{ph}
\end{equation}
This $X$ is also called a Hamiltonian
vector field with respect to the Hamiltonian operator $P$, and $H$
is called the Hamiltonian of $X$.
\end{dfn}

\begin{prp}
For any evolutionary PDE $X\in\B$ and differential operator $P\in\D$,
there exists a unique differential operator $Q\in\D$ such that for any local functionals $F, G$ we have
\[\{F, G\}_Q=X.\{F, G\}_P-\{X.F, G\}_P-\{F, X.G\}_P,\]
The operator $Q$ is called the Lie derivative of $P$ w.r.t. $X$, and
is denoted by $\Lie_XP$.
\end{prp}
\begin{prf}
In fact, suppose $P=\sum\limits_{s=0}^\infty P_s\p_x^s$, we have
\begin{equation}
Q=\sum_{s=0}^\infty\hat{X}(P_s)\cdot\p_x^s-\sum_{k=0}^\infty\left(\frac{\p X}{\p u_k}\cdot\p_x^k\cdot P
+P \cdot (-\p_x)^k\cdot \frac{\p X}{\p u_k}\right), \label{xp}
\end{equation}
where $\cdot$ is the multiplication in $\D$. The proposition is proved.
\end{prf}

The elements of $\LL, \A, \D$ are examples of local $0$-, $1$-, $2$- vectors. In general, we can define
the space of local $k$-vectors $\LL^k_{loc}$ and the Schouten-Nijenhuis bracket between them
\[[\cdot, \cdot]:\LL^k_{loc} \times \LL^l_{loc}\to \LL^{k+l-1}_{loc}.\]
The formulae \eqref{xf}, \eqref{ph}, \eqref{xp} are just particular cases of the Schouten-Nijenhuis bracket, in fact we have
the following correspondence:
\[X.F=[X,F],\quad P.H=[P,H],\quad \Lie_X P=[X, P].\]
The Schouten-Nijenhuis bracket satisfies
the following identities
\begin{align}
&[P,Q]=(-1)^{pq}[Q,P], \label{sb-1}\\
&(-1)^{pr}[[P,Q],R]+(-1)^{qp}[[Q,R],P]+(-1)^{rq}[[R,P],Q]=0, \label{sb-2}
\end{align}
where $P\in \LL^p_{loc},\ Q\in \LL^q_{loc},\ R\in \LL^r_{loc}$.
We omit the general definition of $\LL^k_{loc}$ and the Schouten-Nijenhuis bracket here,
for more details one can refer to \cite{DZ, LZ1, DLZ}.

\begin{dfn}
A Hamiltonian operator $P$ is called of hydrodynamic type if
\begin{equation}\label{zh-5}
P=g(u)\p_x+\frac12g'(u)u_x,\ g(u)\ne0.
\end{equation}
Hamiltonian operators with hydrodynamic leading terms are called generic.
\end{dfn}

\begin{thm}\cite{ddd, Ge}\label{tpc}
Let $P$ be a Hamiltonian operator of hydrodynamic type.
\begin{itemize}
\item[(a)] If $Q\in\D$ is an infinitesimal deformation of $P$, then there exists
$X\in\B$ such that $Q=\Lie_XP$;
\item[(b)] If $X\in\B$ satisfies $\Lie_XP=0$, then there exists a local functional $H\in\LL$ such that $X=P.H$.
\end{itemize}
\end{thm}
This theorem is implied by the vanishing of the Poisson cohomologies \cite{lichn} of the Hamiltonian structure
\eqref{zh-5} as proved in \cite{Ge}(c.f. \cite{sh}), see also \cite{ddd, DZ}.

The identity \eqref{sb-2} implies that for any $X\in\B$, the map
\[g_X:\LL^*_{loc}\to\LL^*_{loc},\ K\mapsto e^{\ad_X}K=\sum_{n=0}^\infty \frac1{n!}\left(\ad_X\right)^n K\]
is an automorphism of the graded Lie algebra $(\LL^*_{loc}, [\cdot,\cdot])$.
Such an automorphism is called a Miura type transformation.

\begin{cor}\cite{ddd, DZ, Ge}\label{cor-2-6}
For any generic Hamiltonian operator $P$ with leading term $P_1$,
there exists a Miura type transformation that reduces $P$ to its leading term $P_1$. In other words,
there exist an $X\in\B$ such that $g_X(P)=P_1$.
\end{cor}

\begin{lem}\label{v22}
Let $X\in\B$ be a Hamiltonian vector field with respect to a generic Hamiltonian operator $P$,
and
\[\e^{-1} X=f(u)u_x+\e X_2+\e^2 X_3+\cdots,\ P=P_1+\e P_2+\e^2 P_3+\cdots\]
then the coefficients $X_2$ and $P_2$ of $\e$ must vanish.
\end{lem}
\begin{prf}
By using Theorem \ref{tpc} we know the existence of $Y=h(u)u_x\in\B$ such that $P_2=\Lie_Y P_1$,
then by using the formula \eqref{xp} we easily obtain $P_2=0$.
Let $H=H_0+\e H_1+\cdots$ be the Hamiltonian of $X$. From the definition of local functionals we see
that $H_1$ always vanishes, so $X_2=[P_1, H_1]+[P_2,H_0]=0$.
The lemma is proved.
\end{prf}

Now let us consider the quasi-triviality
of evolutionary PDEs. Define
\[\tilde{\R}=(\R[u_1^{-1}])_{\ge0}=C^\infty(u_0)[u_1, \frac{u_2}{u_1^2}, \frac{u_3}{u_1^3}, \cdots],\]
whose elements are called {\em almost differential polynomials}
\cite{DLZ}. We can modify the above definition of the spaces $\A$,
$\B$, $\LL$, $\D$ by replacing $\R$ with
$\tilde{\R}$, and denote the resulting
spaces by $\tilde{\A}$, $\tilde{\B}$, $\tilde{\LL}$, $\tilde{\D}$.
The Schouten-Nijenhuis brackets \eqref{xf}, \eqref{ph}, \eqref{xp}
are defined in the same way.

\begin{lem}\label{wqt}
Let $X\in\B$ be an evolutionary PDE of the following form
\begin{equation}
X=\e f(u)u_x+\sum_{i\ge3}\e^i X_i,\ X_i\in\R,\ \deg X_i=i,
\end{equation}
then there exists a unique $Y\in\tilde{B}$ of the form
\begin{equation}\label{quasiy} Y=\sum_{i\ge2}\e^i Y_i,\ Y_i\in\tilde{\R},\ \deg Y_i=i,
\end{equation}
such that $g_Y(X)=\e f(u)u_x$. Any other reducing transformation
$g_Z\,(Z\in\tilde{\B})$ of $X$ can be
represented as $g_{\tilde Y}\circ g_Y$, where $\tilde{Y}=\e p(u)
u_x$ and $p(u)$ is an arbitrary smooth function.
\end{lem}
\begin{prf}
This is a weaker version of the quasi-triviality theorem. The proof is similar to the full version, see \cite{LZ2}.
\end{prf}

\begin{lem} \label{mal}
Let $X=\e f(u)u_x$, $f'(u)\ne0$. If a homogeneous differential operator $P\in\tilde{\D}$ with $\deg P\ge2$ satisfies
$\Lie_XP=0$, then $P=0$.
\end{lem}
\begin{prf}
We only need to consider the case when $f(u)=u$, since other cases can be converted to this case
by applying a change of coordinate $\tilde{u}=f(u)$.
Let the highest order of $\p_x$ in $P$ be $m$, then
\begin{equation}\label{zh1}
P=P_m\p_x^m+\cdots,\ P_m\ne0,\ \deg P_m=\deg P-m\ge2-m.
\end{equation}
The vanishing of the coefficient of $\p_x^m$ in $\Lie_XP$ implies
\[\sum_{k=0}^N \left(\p_x^k(u\,u_1)-u\,u_{k+1}\right)\frac{\p P_m}{\p u_k}+(m-1)u_1 P_m=0.\]
Here $N$ is the biggest indices of $u_k$ that $P_m$ depends on.
By using Lemma 4.1 of \cite{LZ2}, the general solution to the above equation reads
\[P_m=u_x^{1-m} c(u, \phi_2, \phi_3, \cdots,\phi_N),\]
where $c$ is an arbitrary smooth function, and
\[\phi_1=\frac1{u_1},\ \phi_{k+1}=\frac1{u_1}\sum_{i=1}^k u_{i+1}\frac{\p \phi_k}{\p u_i},\ k=1, 2, \cdots.\]
Note that $P_m$ depends polynomially on $u_2, u_3, \cdots$, it follows from the above definition
of the functions $\phi_k$ that
$c$ also depends polynomially on $\phi_2$,$\dots$,$\phi_N$. From the fact that $\deg \phi_k=-1$
we obtain $\deg P_m=1-m+\deg c\le 1-m$,
this contradicts with \eqref{zh1}, so we must have $P=0$. The lemma is proved.
\end{prf}

\begin{cor}
The Hamiltonian operator of a generic evolutionary PDE is also generic.
\end{cor}
\begin{prf}
Let $X=\e f(u)u_x+\cdots$ be a generic evolution PDE, $P=\e^{m-1}P_m+\cdots$, $P_m\ne 0$
be a Hamiltonian operator of $X$,
then we have $\Lie_{f(u)u_x}P_m=0$. If $m\ge2$, then by the above lemma we have $P_m=0$, which
contradicts with our assumption $P_m\ne 0$, so we have $m=1$. The corollary is proved.
\end{prf}

\begin{lem} \label{vh}
Let $P$ be a Hamiltonian operator of hydrodynamic type,
$H\in\tilde{\LL}$ be homogeneous and $\deg
H\ge1$. If $P.H=0$, then $H=0$.
\end{lem}
\begin{prf}
We denote the density of $H$ by $h=F(u, u_x, \cdots, u_N),\ N\ge1$.
Let $X=P.H$, then we have
\[\frac{\p X}{\p u_{2N+1}}=(-1)^N g(u) \frac{\p^2 F}{{\p u_N}^2}=0.\]
So we have $h=F_1(u, u_x, \cdots, u_{N-1})u_N+F_0(u, u_x, \cdots, u_{N-1})$.
After integrating by part, we can assume that the density of $H$ does not depend on $u_N$.
By induction on $N$, we know that the density of $H$ can be chosen to depend only on $u$.
However, by assumption we have $\deg H\ge 1$, so we must have $H=0$. The lemma is proved.
\end{prf}
\vskip 1ex
\noindent {\it Proof of the Main Theorem}:
Let $X\in\B$ be a generic evoutionary PDE
\[X=\e X_1+\e^2 X_2+\cdots,\quad X_1=f(u) u_x,\ f'(u)\ne 0\]
with a Hamiltonian structure $X=P.H$, where
\[P=P_1+\e P_2+\cdots\in D,\quad  H=H_0+\e H_1+\cdots\in\LL.\]
It follows from Lemma \ref{v22} and \ref{wqt} that there exists $Y\in\tilde{\B}$ such that
\[g_Y(X)=\e X_1=\e f(u)u_x.\]
Since $g_Y$ is an automorphim, we have
\[\Lie_{\e f(u)u_x}g_Y(P)=g_Y(\Lie_X P)=0.\]
By using Lemma \ref{mal} we arrive at $g_Y(P)=P_1$.

By using the fact that $g_Y$ is an automorphim again, we get
\[\e f(u)u_x=g_Y([P,H])=[P_1, g_Y(H)],\]
so it follows from Lemma \ref{vh} that $g_Y(H)=H_0$. The theorem is
proved. \hfill $\square$

\section{The Hamiltonian perturbations}

In this section we propose a way to determine whether a given
generic evolutionary PDE is Hamiltonian, and if so, how to find its
Hamiltonian structures.

Lemma \ref{v22} shows that a generic Hamiltonian PDE in $\B$ must take the form
\begin{equation}\label{genx}
X=\e f(u)u_x+\sum_{i\ge3}\e^i X_i,\quad f'(u)\neq 0,\  X_i\in\R,\ \deg X_i=i.
\end{equation}
Moreover, the associated Hamiltonian structure $P$ must be generic, i.e. its leading
term has the form \eqref{p0} with $\phi(u)\neq 0$. According to the Main Theorem,
the reducing transformation $g_Y$ satisfying $g_Y(X)=\e f(u) u_x$ also
reduces $P$ to its leading term $P_1$, where
$Y$ is of the form \eqref{quasiy} given in Lemma \ref{wqt}.
Thus we have
\begin{equation}\label{p}
P=g_Y^{-1}P_1=e^{-\ad_Y}\left( \phi(u)\p_x+\frac12\phi'(u)u_x
\right)
\end{equation}
and the following corollary of the Main Theorem:
\begin{cor} A given generic PDE of the form \eqref{genx} possesses a
Hamiltonian structure of the form \eqref{ham} if and only if there exists a non-vanishing smooth
function $\phi(u)$ such that the bivector \eqref{p} lies in $\D$.
\end{cor}

Given a generic $X$ of the form \eqref{genx}, its reducing transformation $g_Y$ can be found
explicitly by using the method given in \cite{LZ2}. So the right hand side of
\eqref{p} can be computed straightforwardly. In this way we
obtain a criterion for testing whether a generic PDE is Hamiltonian
or not. This algorithm is not hard to be carried out with the help
of a computer program, and we give some examples below to illustrate it.

\begin{emp}\label{ep1}
We consider generic evolutionary PDEs of the form
\begin{equation}\label{eqord2}
u_t=u u_x+\e^2\left(b(u) u_{xxx}+c(u) u_{xx} u_x+d(u) u_x^3\right),
\end{equation}
where $b(u), c(u), d(u)$ are smooth functions of $u$, and $b(u)\ne0$. We are to determine
under what conditions such a system is Hamiltonian.
\end{emp}

\begin{lem}
If the equation \eqref{eqord2} possesses a Hamiltonian structure with leading term \eqref{p0}, then the
functions $b(u), c(u), d(u)$ must satisfy
\begin{equation}\label{abcd}
b^2\,\phi''- 5\,b\,b'\,\phi'+\left(\frac43c^2-16\,b\,d-\frac{20}3c\,b'+8\,{b'}^2+6\,b\,c'-4\,b\,b''\right)\phi=0.
\end{equation}
\end{lem}
\begin{prf}
The reducing transformation of the equation \eqref{eqord2} can be
taken to depend only on even powers of
$\e$. Thus the Hamiltonian structure has the form
$P=g_Y^{-1}P_1=P_1+\e^2 Z_1+\e^4 Z_2+\e^6 Z_3+\e^8 Z_4+\cdots$. A
direct calculation shows that the term $Z_1$ is always a polynomial,
and the polynomiality of $Z_2$ implies \eqref{abcd}.
\end{prf}

The polynomiality of $Z_3, Z_4, \cdots$ leads to a set of ODEs of
$b(u), c(u), \phi(u)$. These ODEs are too complicated to solve. Here
we add a homogeneity
condition to simplify
the problem (see \cite{LZ2} for a similar condition). Since
$b(u)\ne0, \phi(u)\ne0$, without loss of generality we can assume
that
\begin{equation}\label{bc}
b(u)=u^{1-2\alpha}, \ c(u)=\beta u^{-2\alpha},\ \phi(u)=u^\gamma
\end{equation}
where $\alpha, \beta, \gamma\in\mathbb{R}$.
It follows immediately from equation \eqref{abcd} that
\begin{equation}\label{d}
d(u)=\frac{u^{-1-2 \alpha }}{48} \left( 24-72 \alpha +48 {\alpha }^2
- 20 \beta +4 \alpha  \beta +4 {\beta }^2-18 \gamma  +
      30 \alpha  \gamma +3 {\gamma }^2 \right).
\end{equation}

\begin{prp}
The equation \eqref{eqord2} with $b(u), c(u), d(u)$ given by
\eqref{bc}, \eqref{d} possesses a Hamiltonian structure if and only
if the constants $\alpha, \beta, \gamma$ satisfy one of the
following conditions:
\begin{align}
i) & \quad 8\alpha+2\beta+3\gamma-4=0,\label{c1}\\
ii) & \quad \alpha=0,\, 2\beta+3\gamma-6=0, \label{c2}\\
iii)& \quad \beta=0,\, 2\alpha+\gamma-2=0, \label{c3}\\
iv) & \quad(\alpha,\beta,\gamma)=\left(\frac12,-\frac32,0\right),\
\left(\frac12,-3,0\right),\ \left(\frac12,-3,1\right),\
\left(0,-1,\frac{10}{3}\right).\label{c4}
\end{align}
\end{prp}
\begin{prf}
By solving a system of algebraic equations, we can obtain conditions i)-iv) from the polynomiality of $Z_3, Z_4$.
So these conditions are necessary.

Now denote by $\e^{-1} X_1, \e^{-1} X_2, \e^{-1} X_3$ the right hand side of \eqref{eqord2}
such that $\alpha, \beta, \gamma$ satisfy the condition i), ii), iii) respectively.
We define the following differential operators:
\begin{align}
P_1&=u^{\frac{\gamma}2}\,\p_x\,u^{\frac{\gamma}2},\label{pp1}\\
P_2&=u^{\frac{\gamma}2}\left(\p_x+\e^2\p_x^3\right)u^{\frac{\gamma}2},\label{pp2}\\
P_3&=u^{1-2\alpha}\left(u^{2\alpha}\p_x+\alpha\,u^{2\alpha-1}u_x+(\alpha+1)\e^2\p_x^3\right)u^{1-2\alpha}. \label{pp3}
\end{align}
Then by a straightforward computation, we can show that $P_i$ are Hamiltonian operators and $\Lie_{X_i}P_i=0$,
i.e. $P_i$ is a Hamiltonian structure for $X_i$.

In the other four exceptional cases, the equation \eqref{eqord2}
possesses bihamiltonian structures, as it will be shown by the next proposition.
The proposition is proved.
\end{prf}

An evolutionary PDE of the form \eqref{pde} that is given by a
vector field $X\in\B$ is called bihamiltonian, or a bihamiltonian
perturbation of \eqref{pde0}, if there are two compatible
Hamiltonian operators $P, Q$ and two local functionals $H_1, H_2$
such that $X=P.H_1=Q.H_2$. Here the word ``compatible'' means that any
linear combination of $P , Q$ is also a
Hamiltonian operator.

From the Main Theorem it follows that the reducing transformation of the bihamiltonian equation reduces both
Hamiltonian operators $P$ and $Q$ to their leading terms which have the form \eqref{zh-5}. It is easy to see that any two
Hamiltonian operators of the form \eqref{zh-5} are compatible, so the two Hamiltonian operators $P, Q$ of a
bihamiltonian equation \eqref{pde} are automatically compatible.

\begin{prp}
The equation \eqref{eqord2} with $b(u), c(u), d(u)$ given by \eqref{bc},
\eqref{d} possesses a bihamiltonian structure if and only if
the constants $(\alpha, \beta)$ are given by one of the following pairs:
\begin{equation}\label{zh-3}
\left(\frac12,0\right),\ \left(\frac12,-\frac32\right),\ \left(\frac12,-3\right),\
\left(0,-2\right),\ \left(0,-1\right).
\end{equation}
\end{prp}
\begin{prf}
For an equation \eqref{eqord2} determined by the parameters $\alpha,
\beta$, the bihamiltonian condition requires that there exist two
distinct $\gamma_1, \gamma_2$ such that
each $(\alpha, \beta,\gamma_1)$, $(\alpha,
\beta,\gamma_2)$ satisfies one of the
conditions \eqref{c1}--\eqref{c4}. It's easy to check
the pairs of $\alpha, \beta$ given in \eqref{zh-3} are the only ones
that meet the above requirement. Note that for all these cases, the
equation \eqref{eqord2} has been analysed in \cite{LZ2}. The first
three cases are equivalent to the KdV equation via Miura-type
transformations, so they possess bihamiltonian structures which are
induced from that of the KdV equation via the Miura-type
transformations.

For the case when $(\alpha, \beta)=(0,-2)$, we have $\gamma=\frac83, \frac{10}3$, which are implied by the conditions
 \eqref{c1} and \eqref{c2}.
So the corresponding equation has the bihamiltonian structure given by \eqref{pp1} and \eqref{pp2}.

For the case when $(\alpha, \beta)=(0,-1)$, we have $\gamma=\frac83, \frac{10}3$. The first Hamiltonian structure
is given by \eqref{pp2}, and the second one is given by $P_2=u^{\frac43}\,Q\,u^{\frac43}$, where the
differential operator $Q$ reads
\begin{align*}
Q=&w^2\p_x+w w_1\\
&+\e^2\left(2 w^2 \p_x^3+6 w w_1 \p_x^2+4 w_1^2 \p_x+4 w w_2 \p_x+3 w_1 w_2+w w_3\right)\\
&+\e^4\left(w^2 \p_x^5+5 w w_1 \p_x^4+4 w_1^2 \p_x^3+4 w w_2 \p_x^3+3 w_1 w_2 \p_x^2+w w_3 \p_x^2\right),
\end{align*}
and $w=u^{\frac13}$. The proposition is proved.
\end{prf}

\begin{emp} \label{ep2}
Consider the equation
\begin{equation}\label{eqch}
u_t-\e^2 u_{xxt}=u\,u_x+\e^2\left(b\,u\,u_{xxx}+c\,u_{xx}\,u_x\right),
\end{equation}
where $b, c\in\mathbb{R}$. When $b=-\frac13,\ c=-\frac23$ this equation is called the Camassa-Holm equation,
it is an integrable system that
was derived as a shallow water wave equation in \cite{CH1, CH2}, this system also appeared in the study of
hereditary symmetries of soliton equations in \cite{FF}.

Note that if $(b,c)=(-1,-3)$, the equation is equivalent to its leading term $u_t=u\,u_x$.
So we assume $(b,c)\ne(-1,-3)$.
\end{emp}

\begin{lem}
If the equation \eqref{eqch} possesses a Hamiltonian structure with leading term \eqref{p0},
The function $\phi(u)$ must have the form $\phi(u)=\gamma\,u^{d}$.
\end{lem}
\begin{prf}
As in the above example, we can expand the right hand side of \eqref{p} in the form
$P_1+\sum_{k\ge 1} \e^{2 k} Z_k$.
The condition that $Z_2$ and $Z_3$ are differential polynomials yields two ODEs of $\phi(u)$
whose solutions must be power functions.
\end{prf}

Without loss of generality, we assume that $\gamma=1$, i.e. $\phi(u)=u^d$.

\begin{prp}
The equation \eqref{eqch} possesses a Hamiltonian structure if and only if the constants $(b,c,d)$ satisfy one of the following conditions
\begin{align}
i) & \quad d=0,\ c=2b; \label{cc1}\\
ii) & \quad d=0,\ c=3b; \label{cc2}\\
iii) & \quad (b,c,d)=\left(-\frac13,-\frac23,1\right); \label{cc3}\\
iv) & \quad (b,c,d)=\left(0,-1,2\right). \label{cc4}
\end{align}
\end{prp}
\begin{prf}
The proof is similar to that of the previous example. The four Hamiltonian operators under the
coordinate $\tilde{u}=u-\e^2\,u_{xx}$ read
\begin{align}
& P_1=\p_x-\e^2\,\p_x^3; \label{pc1}\\
& P_2=(1+b\,\e^2\p_x^2)(\p_x-\e^2\,\p_x^3); \label{pc2}\\
& P_3=\tilde{u}\,\p_x+\frac12\tilde{u}_x; \label{pc3}\\
& P_4=\tilde{u}\,\left(\sum_{k=0}^\infty \e^{2k}\p_x^{2k+1}\right)\tilde{u}. \label{pc4}
\end{align}
The proposition is proved.
\end{prf}

\begin{cor}
The Camassa-Holm equation is the only
one of the form \eqref{eqch} that possesses a bihamiltonian structure.
\end{cor}

\begin{rmk}
By using the method introduced in \cite{LZ2}, we can prove that
the equation \eqref{eqch} is formally integrable
 if and only if the constants $(b,c)$ is given by
one of the following pairs:
\begin{align*}
&\left(-1,-2\right),\ \left(-1,\frac12\right),\ \left(-1,-\frac12\right),\ \left(-\frac13,-1\right),\\
&\left(-\frac13,-\frac23\right),\ \left(-\frac14,-\frac34\right),\ \left(-\frac14,-\frac12\right).
\end{align*}
Here formal integrability of an evolutionary PDE of the form \eqref{pde}
means, in the sense of \cite{LZ2}, that any symmetry of the equation \eqref{pde0}
can be
extended to a symmetry of \eqref{pde}. Note that
the pair $\left(-\frac13,-\frac23\right)$ corresponds to the Camassa-Holm equation, which is bihamiltonian;
the pair $\left(-\frac14,-\frac34\right)$ corresponds to the Degasperis-Procesi equation \cite{DP,DHH}
which is formally integrable but has only one local Hamiltonian structure.
\end{rmk}

\begin{emp} According to the Drinfeld - Sokolov construction \cite{DS}, The Kac-Moody algebra $A^{(2)}_2$ is
associated to the following evolutionary PDEs:
\begin{align*}
u_{t_1}&=20 u^2 u_x+\e^2\left(10 u u_{xxx}+25 u_{xx} u_x\right)+\e^4
u_{xxxxx}, \\
u_{t_2}&=\frac54 u^2 u_x+\e^2\left(\frac52 u u_{xxx}+\frac52 u_{xx} u_x\right)+\e^4
u_{xxxxx}.
\end{align*}
They have the Hamiltonian structure
$$
\frac{\p u}{\p t_i}=(\e^2 \p^3+2 u \p+ u_x)\frac{\delta H_i}{\delta u},\quad i=1,2.
$$
Here the Hamiltonian $H$ are given respectively by
$$
H_1=\int (\frac43 u^3-\frac12 u_x^2)dx,\quad H_2=\int (\frac1{12} u^3-\frac12 u_x^2)dx.
$$
The above algorithm can be carried out to show that the above two equations have no other
local Hamiltonian structures.
\end{emp}

\section{Concluding remarks}

We showed that the reducing transformation of a
Hamiltonian perturbation \eqref{pde} of the hyperbolic equation
\eqref{pde0} also reduces its Hamiltonian structures to
their leading terms. We also applied this fact to the
study of existence of Hamiltonian structures for a generic
evolutionary PDE of the form \eqref{pde}. Note that the derivation
of the these results relies on the fact that the evolutionary PDEs
we are considering are scalar.
It is interesting to generalize these results to evolutionary PDEs that are perturbations of
certain multi-component hyperbolic systems, such generalization is important in particular in the study of
integrability for certain nonlinear evolutionary PDEs. We finish the paper with the following remarks:

\begin{rmk}
Let us note that the bihamiltonian equations which appear in Examples \ref{ep1} and \ref{ep2}
are also formally integrable in the sense of \cite{LZ2}, i.e. any symmetry of their leading terms can be
extended to a symmetry of themselves. By combining the main results of \cite{LZ1}, \cite{LZ2} and the present paper,
we can prove this fact for any evolutionary PDEs of the form \eqref{pde}.
Furthermore, the main results of \cite{DLZ} and \cite{LZ1} imply the following theorem:
\begin{thm}
Let $(P, Q)$ be a semisimple bihamiltonian structure of hydrodynamic type with leading term $(P_1, Q_1)$.
Then any bihamiltonian vector field of $(P_1, Q_1)$ can be extended to a bihamiltonian vector field of $(P, Q)$.
\end{thm}
We note that the semisimplicity condition is automatically satisfied in the scalar case.
\end{rmk}

\begin{rmk}
The quasi-Miura transformation \eqref{qt1} that reduces the equation \eqref{pde} to its leading term \eqref{pde0}
is understood as a formal power series in $\e$. When \eqref{pde} is a Hamiltonian perturbation, it is conjectured
by Dubrovin in
\cite{Du} that for any smooth solution $v(x,t)$ to \eqref{pde0} defined for all $x\in\mathbb{R}$ and $0\le t<t_0$
which is monotone in $x$ for any $t$, the associated reducing transformation gives the
asymptotic expansion at $\e\to 0$ of a solution $u=u(x,t;\e)$ to \eqref{pde} defined on the same domain in the
$(x,t)$-plane.

Note that for some nonlinear PDEs that are covariant under the action of
certain Lie groups, Flato and
his collaborators developed a program of linearization of such equations, they constructed transformations on
appropriate Banach or
Fr\'echet  spaces which reduce the nonlinear
PDEs to the equations given by their linear parts, and in this way solved the associated Cauchy problems,
see \cite{F1, F2, F3} and references therein. The reducing transformations considered in this paper
may be viewed as the quasi-linear counterpart of the linearization transformations. We hope that the
analytic method of the program of linearization may give us a hint to the study of the above mentioned
conjecture of Dubrovin.
\end{rmk}

\vskip 0.4truecm \noindent{\bf Acknowledgments.}
The authors thank Boris Dubrovin and Daniel Sternheimer for helpful discussions and
suggestive comments on the subject of the paper.
This work is partially supported by NSFC No.10631050 and the
National Basic Research Program of China (973 Program) No.2007CB814800.

\end{document}